\documentclass[showpacs,preprint,amsmath,superscriptaddress,prl]{revtex4}  
\usepackage[dvips]{graphicx}
\usepackage{epsfig}
\usepackage{color}
\usepackage{subfigure}
\usepackage{soul}

\begin{document} 
\title{
Truncated
lognormal distributions and scaling
in the size of naturally defined population clusters\\
} 
\author{\'Alvaro Corral}
\affiliation{%
Centre de Recerca Matem\`atica,
Edifici C, Campus Bellaterra,
E-08193 Barcelona, Spain
}\affiliation{Departament de Matem\`atiques,
Facultat de Ci\`encies,
Universitat Aut\`onoma de Barcelona,
E-08193 Barcelona, Spain
}\affiliation{Barcelona Graduate School of Mathematics, 
Edifici C, Campus Bellaterra,
E-08193 Barcelona, Spain
}\affiliation{Complexity Science Hub Vienna,
Josefst\"adter Stra$\beta$e 39,
1080 Vienna,
Austria
}
\author{Frederic Udina}
\affiliation{Department of Economics and Business, Universitat Pompeu Fabra, 
Ramon Trias Fargas 25-27, E-08005, Barcelona, Spain}
\author{Elsa Arcaute}
\affiliation{Centre for Advanced Spatial Analysis, University College London, 90 Tottenham Court Road, London, W1T 4TJ, UK}
\begin{abstract} 
Using population data of high spatial resolution
for a region in the south of Europe,  
we define cities by aggregating individuals to form connected clusters.
The resulting cluster-population distributions show a smooth decreasing behavior
covering six orders of magnitude.
We perform a detailed study of the distributions, using state-of-the-art statistical tools.
By means of scaling analysis we rule out the existence of a power-law regime 
in the low-population range.
The logarithmic-coefficient-of-variation test allows us to establish
that the power-law tail for high population, characteristic of Zipf's law, 
has a rather limited range of applicability.
Instead, lognormal fits describe the population distributions
in a range covering from a few dozen individuals to more than one million
(which corresponds to the population of the largest cluster).

\end{abstract}

\maketitle

\section{Introduction}

Cities are expected to experience enormous grow in the next decades,
but already nowadays, 
they can be considered in some sense as
the largest structures built by humankind.
However, in contrast to other human constructions, 
cities develop from social and economic processes
combined with top-down planning. 
Social and economic processes in their turn 
depend on technological and scientific advances.
So, cities are complex systems driven by self-organization, 
where its fundamental constituents (the individuals)
participate in a vast number of different types
of interactions that keep the city alive \cite{Barthelemy_cities_review}.
In fact, the analogy between cities and living organisms
is not just a metaphor but a very deep insight \cite{Bettencourt_West}.

Probably, the first characterization of any entity is in terms of its size, and this also holds for complex entities.
It is well known that 
for cities, their size 
(measured for instance in number of inhabitants)
is broadly distributed (there are cities of vastly different sizes, 
taking a broad definition of a city as a ``human settlement'').
Then, a statistical description is necessary.
Several statistical models for city size have been proposed, 
with the most important one being
Zipf's law \cite{Zipf_1949},
which states that, given a country or a large region, 
the probability mass function $f(s)$ of city size $s$
is given by a power-law (pl) distribution,
$$
f_{pl}(s) \propto \frac 1 {s^\beta},
$$
with 
the symbol ``$\propto$'' denoting proportionality
and
the exponent $\beta$ taking values close to two
(an important requirement is that the exponent has to be larger than one).
The law should apply at least to the largest cities,
i.e., for the upper tail of the size distribution,
and so one has in mind cities and towns but not necessarily small villages.
It is a remarkable fact that Zipf's law seems to hold in many
other systems in which individuals gather into some sort of groups
or classes (companies \cite{Axtell}, religions \cite{Clauset}), 
and where the ``individuals''
can be anything from animals \cite{Pueyo} 
to links in the Internet \cite{Adamic_Huberman}, 
word tokens in a text \cite{Moreno_Sanchez},
or combinations of musical notes \cite{Serra_scirep}. 

Nevertheless, there have been authors who have argued in favor of other models;
in particular, for city-size distribution
the lognormal model has been proposed as 
the most remarkable alternative to Zipf's law, and some debate has arisen
\cite{Eeckhout,Levy_comment,Malevergne_Sornette_umpu}.
This debate can be put in the broader context of the adequacy
of power-law fitting procedures 
\cite{White,Clauset,Corral_nuclear,Corral_Deluca,Barabasi_criticism,Corral_Gonzalez,Voitalov_krioukov},
but is certainly different from the controversy 
about power-law relations in ``urban metabolism''
or urban allometry \cite{Bettencourt_West,Arcaute_scaling,Leitao}.
Nevertheless, at the core of both problems is the proper use of statistical tools, 
which is part responsible of the recent, unfortunate, and deep problem
known as reproducibility crisis, or replicability crisis
\cite{Peng_reproducibility,ASAstatement}.

In any case, 
one can realize that there is a degree of arbitrariness
in city-statistics research, related to the definition of what a city is.
If the usual administrative delimitations
(which
were established, in general,
following criteria developed
many decades or centuries ago) 
are used for today urban agglomerations, 
to which extend are the results  for city-size distributions not just an artefact of old bureaucracy? 
Clearly, more realistic and scientific definitions of the concept of city are necessary.
This has been attempted by several authors
\cite{Rozenfeld,Jiang2011,Jiang2014},
who have introduced the concept of naturally-defined cities,
see in particular the citations in Refs. 
\cite{Rozenfeld,Arcaute_Britain,Arcaute_scaling}.

In the present paper we use population data of high resolution
to construct clusters of population, which we identify with cities,
whose size distribution is scrutinized with state-of-the art statistical tools.
In the next section we describe the data; 
in Sec. 3 we explain the several similar procedures used to 
construct the population clusters (our definition of cities);
and
in Sec. 4 we present our statistical study of the size of clusters, 
using scaling analysis, the logarithmic-coefficient-of-variation test, 
as well as truncated-lognormal and power-law fits
of the resulting distributions.
We anticipate that the lognormal distribution is much more suitable
than the power law to describe the cities arising from the analyzed dataset
and the city-definition introduced.
Also,
the importance of spatial correlations in the number of inhabitants
to get a broad cluster-size distribution is clearly established.

\section{Data}

Here we approach the problem 
of city definition and the validity of Zipf's law
using high-resolution data 
for the scatter of a population through a territory.
The territory under study is
Catalonia ({\it Catalunya}), located in NE Spain and whose capital is the colorful city of Barcelona.
Catalonia 
has a population of about 7,500,000 inhabitants in a area of 32,000 km$^2$, which 
yields an average density around 230 inhabitants per km$^2$
and classifies Catalonia as a highly populated area.
Note that these figures are similar to those of some small European countries, 
such as Switzerland for example.

In Spain, the municipality councils ({\it ayuntamientos})
collect a population register called {\it Padr\'on Municipal de Habitantes}.
All citizens are required to be registered in some municipality and actually 
it is necessary to be registered to access most of the administrative services 
like health, education, etc.
The coordination of the registers of all municipalities in the country is done by
the Spanish {\it Instituto Nacional de Estad\'{\i}stica}
({\tt http://www.ine.es}), 
which sends the information
referred to Catalonia to the Catalan {\it Institut d'Estad\'{\i}stica de Catalunya}
(IDESCAT, {\tt http://www.idescat.cat}).
The processing of the registers is an  important step because 
it guarantees their high quality: 
duplicated entries are removed as are people deceased or registered in a foreign-country embassy.

In the last years, IDESCAT has undertaken the task of georeferencing each individual's postal address present in the register, by means of the  
geocoding web service of the {\it Institut Cartogr\`afic i Geol\`ogic de Catalunya}
({\tt http://www.icgc.cat}),
which assigns geographical coordinates to each postal address.
The complete procedure including the imputation for missing data
is detailed in Ref. \cite{Sune}.

\begin{figure}[ht]
\includegraphics[width=.45\columnwidth]{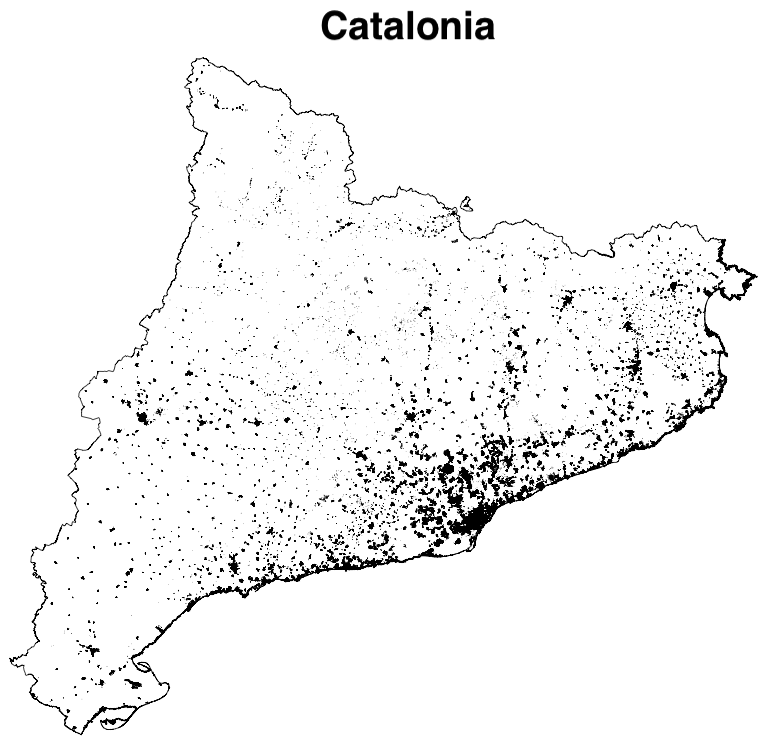}
\includegraphics[width=.43\columnwidth]{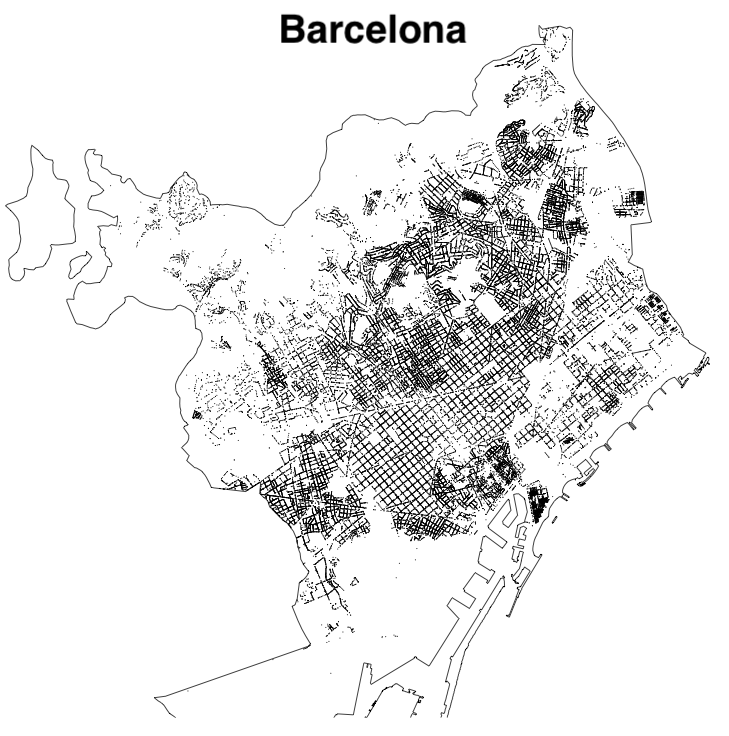}
\caption{
(a) Whole dataset:
Latitude and longitude of the 7,586,888 inhabitants of Catalonia 
at January 1, 2013.
(b) Zoom of the data around the Barcelona zone.
Notice that we are representing the coordinates of the residence place 
of each individual, so, the high resolution of the data becomes apparent
in this plot.
}
\label{Catalonia}
\end{figure}

The data that we have used for our study is
the georeferenced population of Catalonia at January 1, 2013, 
with a total population of $M=$7,586,888 inhabitants
in 989,997 places of residence (i.e., domicile buildings),
and with a 7.6 \% of errors in the georeferencing
for which the procedure of imputation is applied \cite{Sune}.
This register can be considered as high resolution population data, 
even of higher resolution than the data used in Refs. \cite{Orozco,Semecurbe}
(which was 100 and 200 m, respectively;
our is about few meters, corresponding to the minimum distance
between places of residence).
The spatial distribution of the complete data set is displayed in Fig. \ref{Catalonia}.

\section{Clusters of population}
\label{sec3}

\subsection{Grid approach}

In order to construct our aggregations of population, 
we first work using a simple (equirectangular) projection
of longitude and latitude into Cartesian coordinates,
which introduces very little distortion due to the small
extent of the territory. 
We cover the resulting projection by a grid composed by 
identical square cells, each of fixed width $\ell$ in degrees and projected area $\ell \times \ell$,
aligned with the longitude-latitude axes.
In a second, more refined approach, we transform longitude and latitude into distances
(using that $1^\circ$ in latitude is equal to 111.1 km
and
$1^\circ$ in longitude is about 83 km at latitude $41^\circ$), 
and introduce again a square grid.
We call these two approaches grid-in-degrees and grid-in-km, respectively.
We advance that both of them will lead to essentially the same results.

Note that a square in longitude-latitude is equivalent to a rectangle in distance, 
and vice versa, so, in terms of distances our two types of grids are
rectangles (with fixed aspect ratio) and squares, respectively.
When we report the width $\ell$ of a cell in degrees 
it is implicit that we are dealing with the first approach,
and when $\ell$ is in meters or in km we will follow the second one.

\begin{figure}[ht]
\includegraphics[width=.48\columnwidth]{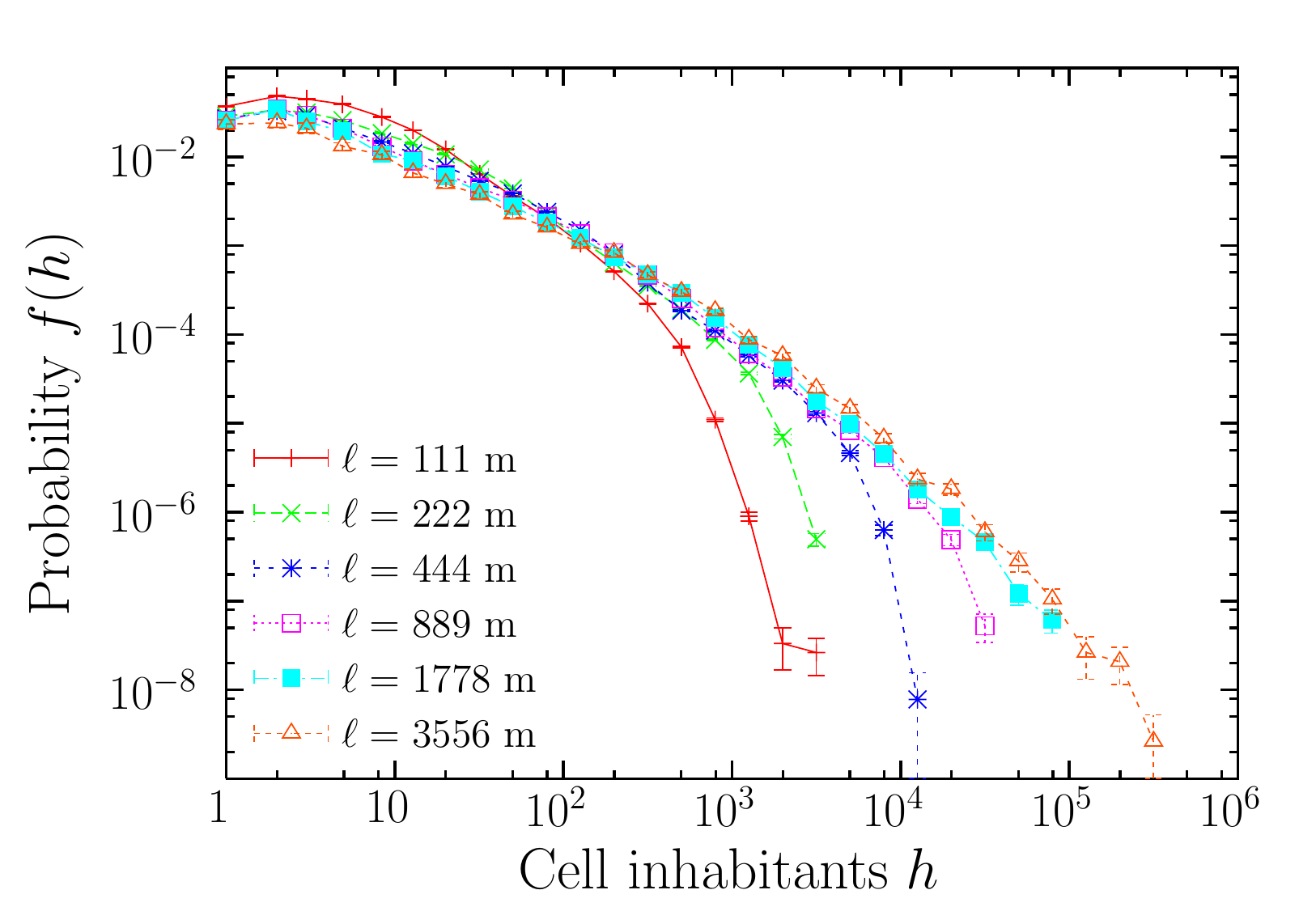}
\includegraphics[width=.48\columnwidth]{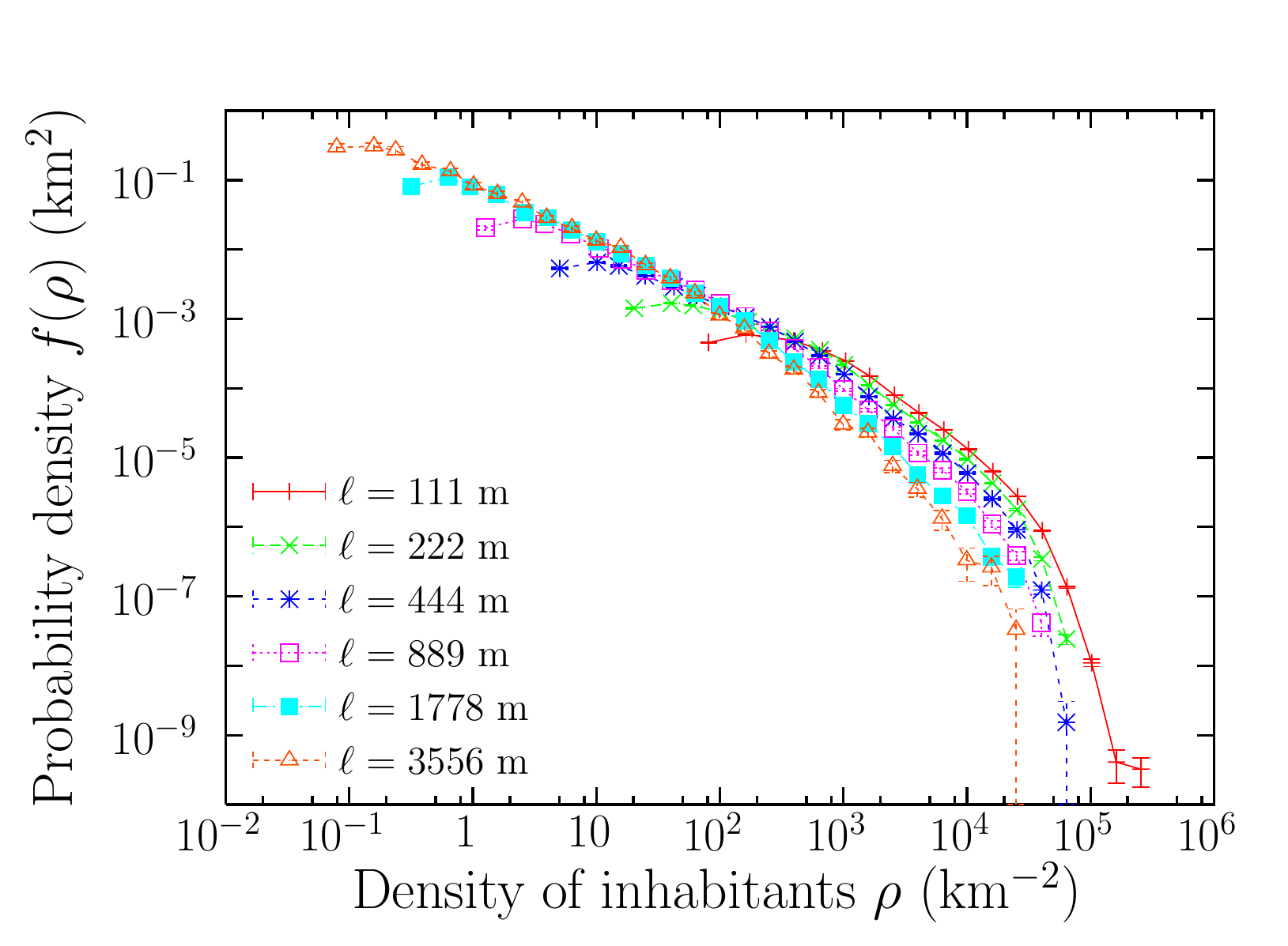}
\caption{
(a) Empirical probability mass functions $f(h)$
of number of inhabitants $h$ per cell, 
for several values of cell width $\ell$, in the grid approach , 
using the grid-in-km procedure. 
(b) Corresponding empirical probability densities of population density per cell.
Observe the enormeous variability, from less that 0.1 inhabitant per km$^2$
to more than $10^5$.
Unpopulated cells are not considered.
}
\label{Dindividualcells}
\end{figure}

The next step is counting the number of inhabitants $h$ in each cell.
For reasonable values of the cell width (for instance, $\ell=0.001^\circ$)
the resulting $h$ turns out to be broadly distributed, from one inhabitant per cell
to many thousands (we will disregard unpopulated cells, 
for reasons that will become clear later). 
For the sake of illustration, we display the corresponding probability mass function  $f(h)$
in Fig.~\ref{Dindividualcells}(a)
for different values of the cell width $\ell$
for the grid-in-km approach,
confirming the broadness of the distribution
(dependence of $f(h)$ on $\ell$ is obviated in the notation).
The population density in each cell can be
calculated straighforwardly as $\rho=h/\ell^2$, 
and its probability density $f(\rho)$ 
(the probability density of the population density)
is shown in Fig.~\ref{Dindividualcells}(b), in units of km$^2$.
It is obvious that, for the same $\ell$, both distributions, $f(h)$ and $f(\rho)$, have the
same shape, with the only difference of the scale factor $\ell^2$. 
The fact that we consider $f(h)$ as a probability mass function 
and $f(\rho)$ as a probability density is not relevant, and 
comes from the consideration 
of $h$ as a discrete variable and $\rho$ as a continuous one, 
but this difference does not carry any deep meaning.

Under the present grid approach, 
our definition of city is based on the aggregation of adjacent
occupied cells. 
This is a natural definition, previously used in Ref. \cite{Rozenfeld}.
We will consider a cell as occupied if its population is greater or equal than a 
threshold value, and unoccupied otherwise.
Naturally, the most immediate value for the population threshold is one \cite{Rozenfeld},
but other prescriptions are possible;
the advantage of our high-resolution data
is that the threshold can be made as small as desired,
in contrast for instance to Ref. \cite{Arcaute_scaling}.
For this reason, the occupation threshold is equal to one in this work.

More concretely,
as in the problem of site percolation \cite{Aharony,Christensen_Moloney}, 
a set of nearest-neighbor occupied cells 
surrounded by non-occupied nearest neighbors defines a (connected) cluster \cite{Rozenfeld}. These clusters
will constitutute a proxy for cities
(we identify the clusters by means of a variation of the classic
Hoshen-Kopelman algorithm);
and we may generically refer to \emph{clusters of population}.
As in this framework the definition of what a cluster (or a city) is depends
on $\ell$, and there is no a-priori way to find an optimum $\ell$,
different values of this parameter will be considered, 
in order to test the robustness of the results.

\begin{figure}[ht]
\includegraphics[width=.90\columnwidth]{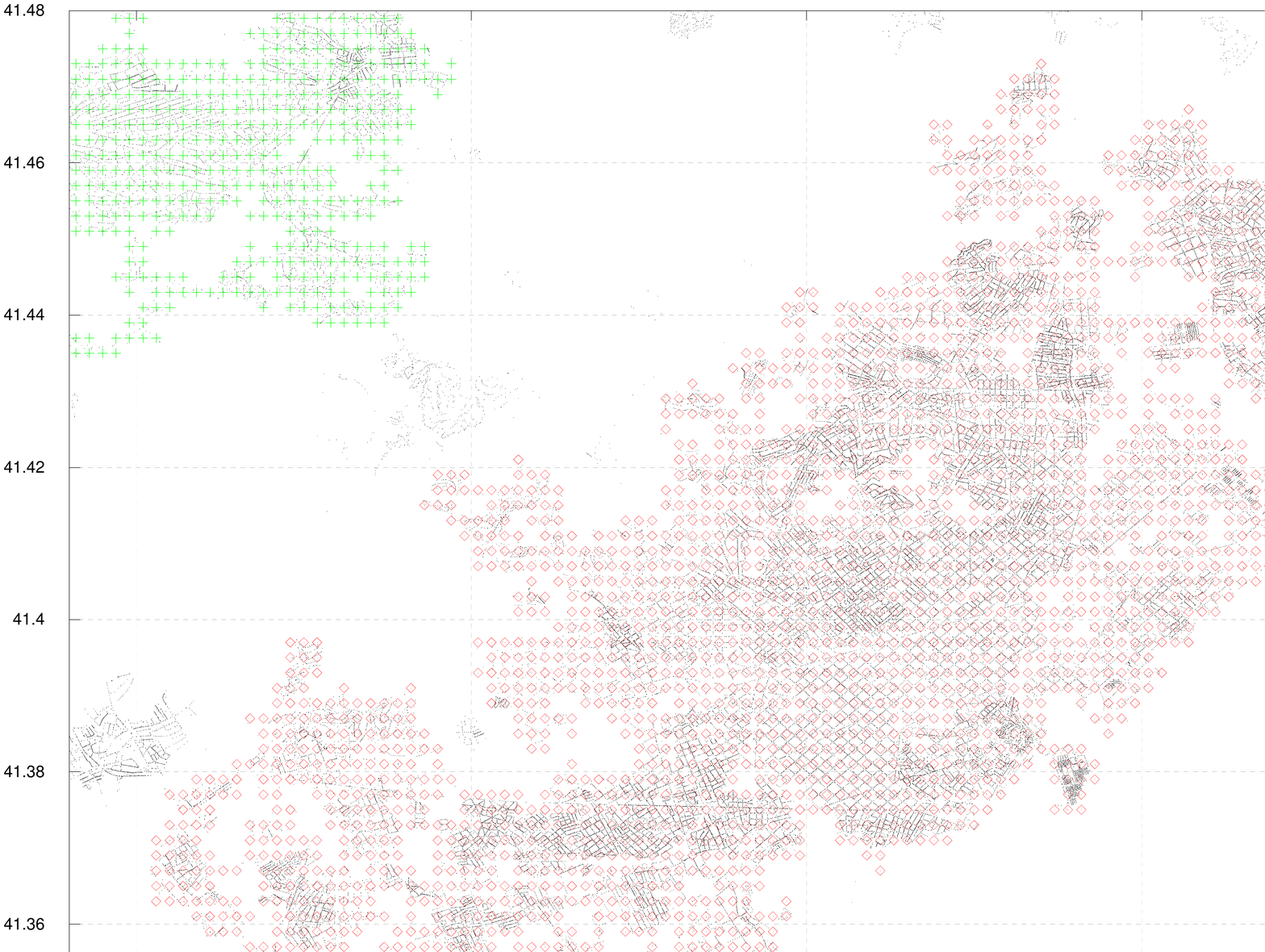}
\caption{
Cluster associated to Barcelona urban area (in red),  for cell width $\ell=0.002^\circ$.
The cluster, which is the largest one for this $\ell$, includes part of other municipalities in addition to Barcelona,
but there are parts of the Barcelona municipality not included in the cluster.
Part of the second-largest cluster (in green) is also shown.
}
\label{bcn}
\end{figure}

The size $s$ of a cluster is defined as its total population
(do not get confused with its total area), 
i.e., for a cluster $i$,
\begin{equation}
s_i = \sum_{\forall  j \in i} h_j,
\label{si_population}
\end{equation}
where the sum runs for all cells $j$ that are part of cluster $i$
(obvioulsy, the cluster definition implies that 
no cell can belong to more than one cluster).
Note that the cluster sizes can range
from 1
to the whole population of the territory
(depending on the spatial location and on
the selected value of the underlying cell width $\ell$).
Then, one should not find strange in this context
to talk about cities with just one inhabitant, 
although it is more proper to refer to them as size-one clusters.
Table \ref{table1new} 
provides, for different values of $\ell$,
the total number of clusters and
the size of the largest one (in terms of numbers of inhabitants)
resulting from applying our procedure.
As an example, Fig. \ref{bcn} shows the largest cluster for cell width $\ell=0.002^\circ$.

\subsection{Ball approach}

An alternative approach to define clusters of population 
can be done using the CCA  
(City Clustering Algorithm \cite{Rozenfeld});
this percolation method has been previously employed by one of the authors, 
see Ref. \cite{Arcaute_Britain}. 
Its implementation can be done using a DBSCAN algorithm
({Density-based spatial clustering of applications with noise \cite{DBSCAN1996}}), 
changing the distance at each iteration.
The approach
 is based on considering ``balls'' of a fixed radius $\ell$,
centered on each individual;
given a radius value $\ell$, a cluster is defined 
as the set of all balls that overlap with (or touch) 
at least another ball in the cluster.
This means that any individual in the cluster
is at a distance smaller than (or equal to) $2\ell$ 
of at least another individual in the cluster.
The distance is measured in meters over the Earth surface.

The cluster population is obtained again as the sum of individuals
contained in the cluster.
One could still use Eq. (\ref{si_population}),
but then $h_i$ has to be interpreted as the number of individuals 
in the center of each ball
and the sum has to run for all balls $j$ associated to cluster $i$.
We refer to this procedure as the ball approach,
and it has clear advantages with respect the grid approach that
it is not affected by the arbitrariness of setting an origin of coordinates for the grid
and that it avoids problems in defining a grid over a sphere
(mainly if one pretends to extend this kind of analysis to much larger regions).
The properties of the clusters resulting from this approach are included in Table \ref{table1new}.
Once the population has been computed for each cluster, 
the cluster-size distribution follows immediately;
this will be shown in the next section.

\section{Analysis and results}

\begin{figure}[ht]
\includegraphics[width=.90\columnwidth]{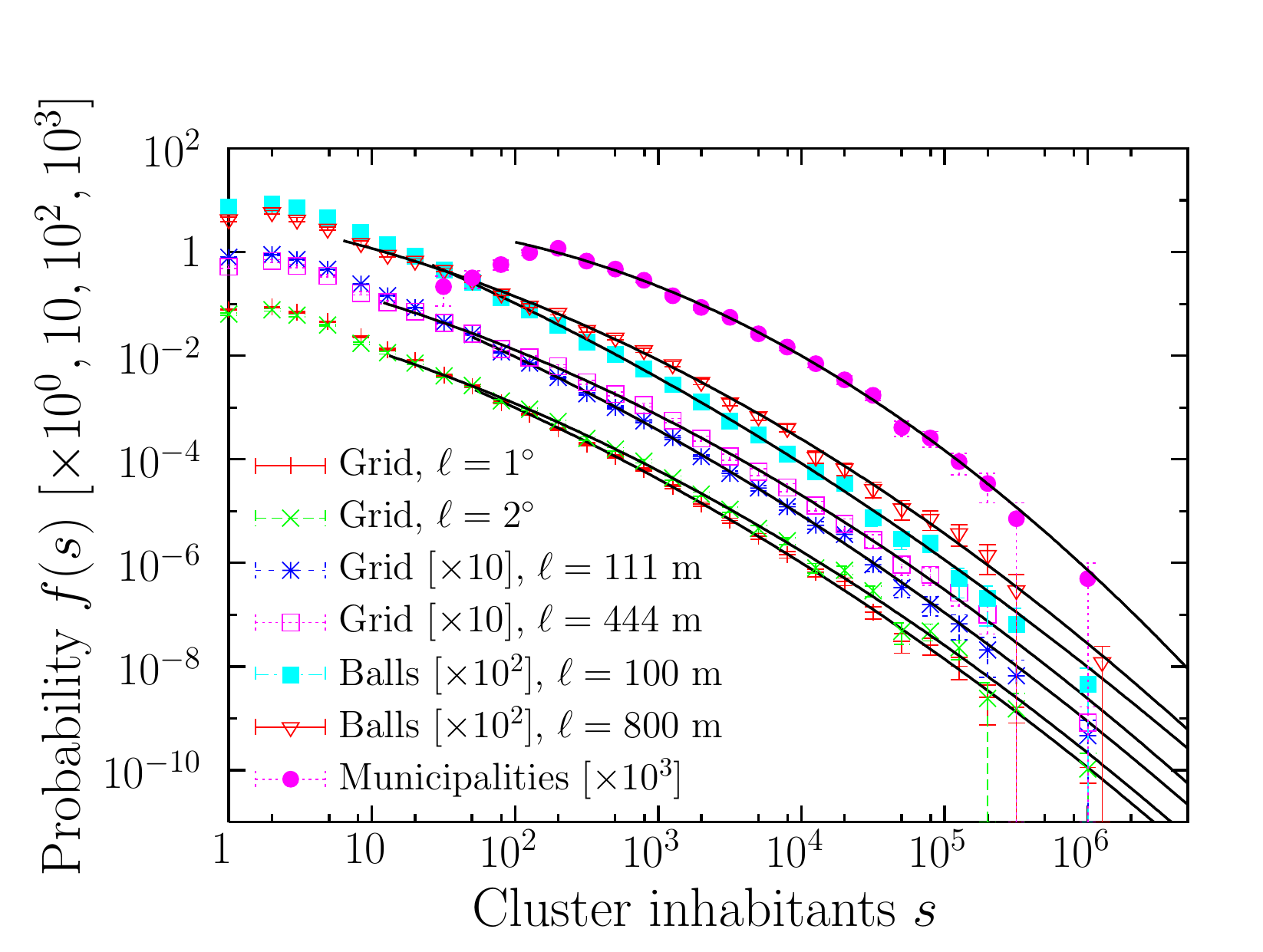}
\caption{
Empirical probability mass functions $f(s)$ of cluster size $s$
(in number of inhabitants) for several values of cell width $\ell$, 
using: 
square grid in longitude-latitude (grid-in-degrees, bottom curves), 
square grid in distance (grid-in-km, $y-$axis multiplied by a factor $10$, 
for clarity sake), 
and ball approach ($y-$axis multiplied by a factor $100$).
Results for the usual approach based on municipalities are also included,
for the sake of comparison.
Continuous lines are lognormal fits.
}
\label{fs}
\end{figure}

In order to investigate the validity of Zipf's law for the clusters of population, 
we consider, as in Ref. \cite{Clauset}, 
the point of view of the distribution of sizes
(in contrast to using the rank-size relation,
which can lead to confusing interpretations \cite{Cristelli}).
The advantages of this choice are discussed in Ref. \cite{Corral_Cancho}, 
see also Ref. \cite{Moreno_Sanchez}.
Figure \ref{fs} displays the corresponding cluster-size distributions 
in terms of the empirical probability mass function $f(s)$,
for different values of $\ell$
(the notation obviates 
the dependence on $\ell$)
and for the three approaches (grid-in-degrees, grid-in-km, and balls).

We clearly observe the broadness of the distributions, 
ranging from population one to more than one million
(more than 6 orders of magnitude).
The smoothness of the distributions is also apparent,
with no change of behavior for all the range, 
except, perhaps, in the transition from one inhabitant to two,
where the probability of former value (one)
is decreased with respect to the latter
(in a sense, one could speculate that a fundamental unit of human population could be the couple, instead of the single individual).
In contrast, the usual distribution of population for the municipalities
(also included in the plot) shows a clear transition around population 200; 
thus, the presence of villages with population below this value is
greatly diminished.
With our definition of population clusters, instead,
the broadness and smoothness of the distributions
do not allow to find discontinuity points to distinguish between 
cities and towns, and between towns and villages
(except for the change of behavior between clusters of size one and two,
as just mentioned).
In addition, the proximity of the distributions to a straight line in 
log-log representation
suggests a power-law behavior.
However, it is misleading to use visual information
of linear behavior in log-log plots as an indication of power-law behavior
\cite{White,Clauset}. 
Rigorous statistical tools are required \cite{Clauset,Corral_Deluca,Corral_Gonzalez,Voitalov_krioukov}.

\begin{figure}[ht]
\includegraphics[width=.48\columnwidth]{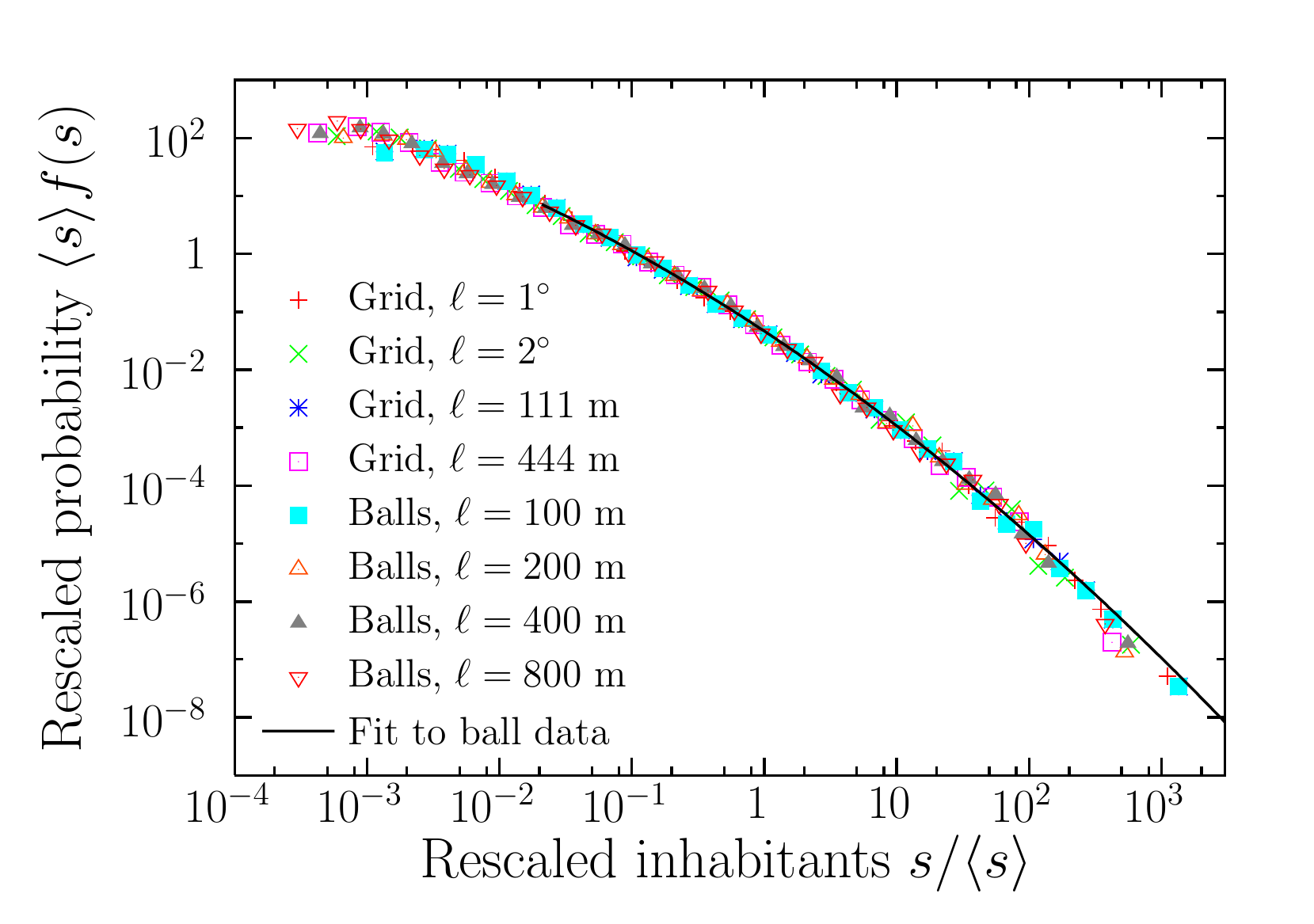}
\includegraphics[width=.48\columnwidth]{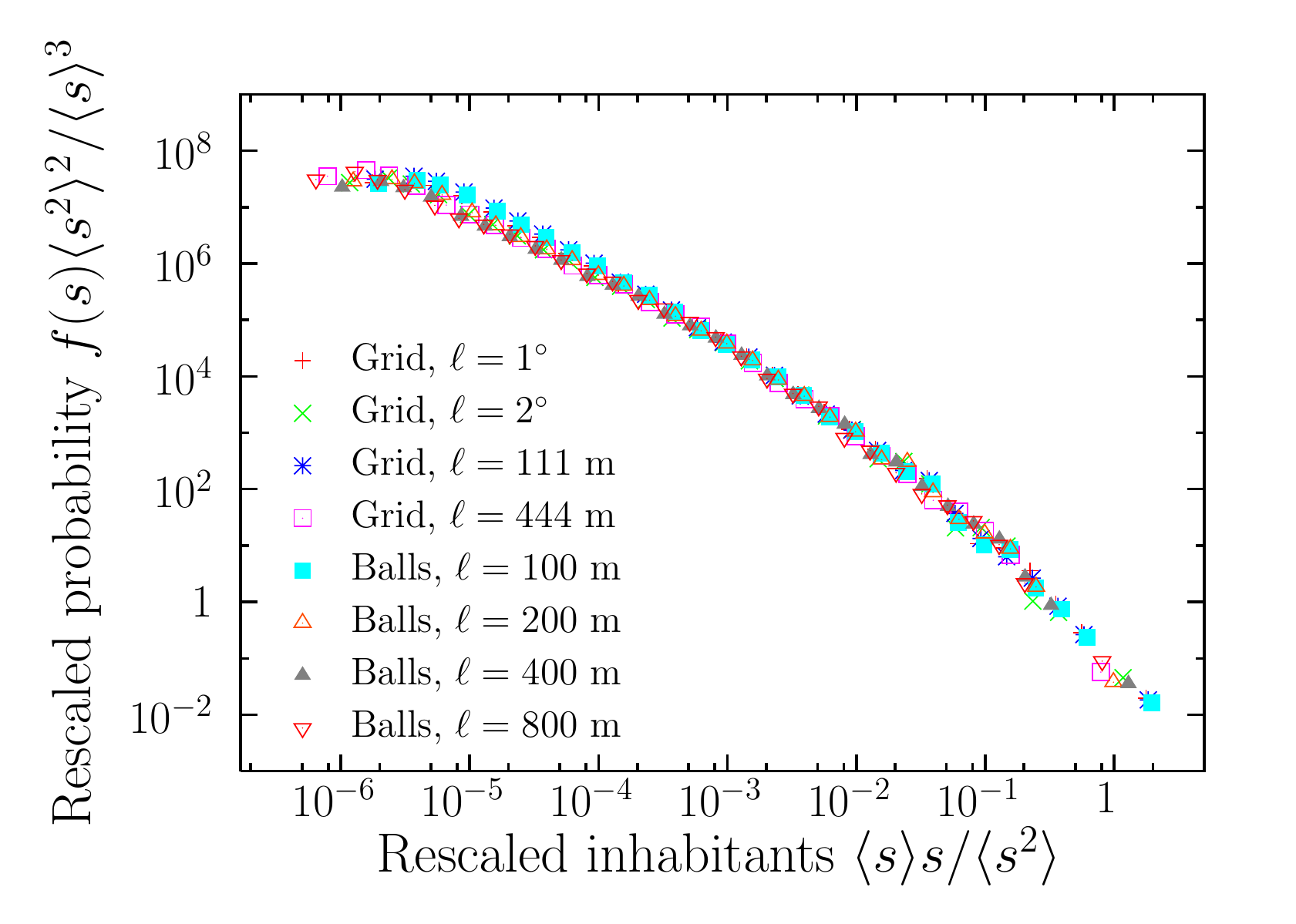}
\caption{
(a)
Empirical probability mass functions $f(s)$
rescaled by their mean value, $\langle s \rangle$
for some values of $\ell$
in the grid-in-degrees, grid-in-km, and ball approaches.
The curve is the fit corresponding to the ball approach, as shown in the table. 
(b) Same probability mass functions using the non-trivial rescaling
given by Eq. (\ref{scalingscaling}), see also Ref. \cite{Corral_csf}.
The scalings in (a) and (b) can be considered as visually equivalent, 
which rules out the existence of a power-law with $\alpha > 1$ for small $s$.
This could be made more quantitative using the collapse algorithm 
of Ref. \cite{Deluca_npg}.
}
\label{fsrescaled}
\end{figure}

\subsection{Scaling analysis}

As a 
first step before moving to more quantitative methods,
we apply scaling analysis to the distributions
(do not get confused between scaling and power-law behavior; the distinction will become clear in what follows).
When one has several broad distributions, which depend on some parameter
($\ell$ in our case), scaling analysis can be a very informative tool \cite{Corral_csf}.
Let us assume that, for different values of the cell width $\ell$, 
the cluster-size distributions $f(s)$ scale with 
some scale parameter $\theta$ (which depends on $\ell$)
as 
$$
f(s)\simeq \frac K \theta G\left (\frac s \theta\right),  
$$ 
where $K$ is a normalization ``constant''
(which could depend on $\theta$)
and $G$ is the scaling function
(which is the same no matter the value of $\theta$, i.e., of $\ell$).
It turns out that when, for small arguments, $G$ does not behave as a power law,
or behaves as power law with exponent smaller than 1, 
the scaling law can be rewritten as
$$
f(s)\propto \frac 1 {\langle s \rangle} G\left(\frac s {\langle s \rangle}\right), 
$$
due to the fact that the mean $\langle s \rangle$ scales linearly with $\theta$
and the constant $K$ is a true constant
($K$ and the constant relating $\langle s \rangle$ to $\theta$ are absorbed into $G$).

However, when for small arguments, 
$G$ diverges as a power law with exponent
$\alpha$ greater than one (but smaller than two), 
the previous scaling law is not valid and
one instead has
\begin{equation}
f(s)\propto \frac 1 \theta \left(\frac m \theta\right)^{\alpha-1}
G\left (\frac{s}{\theta}\right)
\propto \frac {\langle s \rangle^3}{\langle s^2 \rangle^2} 
G\left (\frac{\langle s \rangle s }{\langle s^2 \rangle}\right),
\label{scalingscaling}
\end{equation}
as $\langle s\rangle \propto \theta^{2-\alpha}$ and
$\langle s^2\rangle \propto \theta^{3-\alpha}$,
with $m$ the minimum value of $s$ (below which $f(s)$ is zero)
and $\langle s^2 \rangle$ the second moment of the distribution
(see Ref. \cite{Corral_csf}).
In fact, this new scaling law is also valid in the other case 
($\alpha < 1$ or absence of power law), 
due to the trivial scaling $\langle s\rangle \propto \theta$
and $\langle s^2\rangle \propto \theta^2$ there,
but the reciprocal is not true.
Note that now we refer to the power-law exponent as $\alpha$,
in order to distinguish it from $\beta$, the power-law exponent at the large-$s$ tail.

Therefore, 
estimating the moments from the sample and
plotting, for different values of the cell width,
$\langle s\rangle f(s)$ versus $s/\langle s\rangle$ 
as well as $\langle s^2\rangle^2 f(s) / \langle s\rangle^3$ 
versus $\langle s\rangle s/\langle s^2\rangle$,
one will be able to check
not only if a scaling law holds, 
but if for small arguments
the distribution has a power-law shape with exponent  in the range $1 < \alpha < 2$.
This is done in Fig. \ref{fsrescaled}
for the three representations (grid-in-degrees, grid-in-km, and balls);
the data collapse for all analyzed $\ell$ indicates
that both scaling laws are indeed fulfilled, and
this implies that the power-law behavior (with $\alpha>1$) can be discarded.
The data collapse also shows that for different cell width $\ell$, 
all analyzed cluster-size distributions have (roughly) the same shape, 
but at different characteristic scale, 
and this shape is not a power law (with $\alpha > 1$), at least for small $s$.

\subsection{Residual logarithmic coefficient of variation}

Still it could happen that we had a power law not for small $s$ but for the tail.
In order to investigate this
we apply the test proposed in Ref. \cite{Malevergne_Sornette_umpu}
to compare the performance of a lognormal tail versus a power-law tail.
By tail we mean the part of the distribution that is above an arbitrary threshold value $s_{cv}$
of the random variable;
in other words, the tail is given by the domain $s > s_{cv}$.
We expect $s_{cv}$ to be relatively large.
The test proceeds by computing the (residual) coefficient of variation $cv$ of $\ln (s/s_{cv})$, which is
\begin{equation}
cv = \frac 
1 {\sqrt{n_{cv}-1}} 
\frac{\sqrt{\sum_i \left(\ln s_i - \overline{\ln s}\right)^2}}{\overline{\ln s} - \ln s_{cv}},
\label{CVlog}
\end{equation}
with 
$\overline{\ln s}=n_{cv}^{-1} \sum_i \ln s_i$,
the sums comprising only the values $s_i$ 
above $s_{cv}$ (the ``residual'' values),
and $n_{cv}$ counting the number of data fulfilling this condition
(in practice, $s_{cv}$ is set equal to an empirical value, 
which is excluded then from the tail, due to the strict inequality $s> s_{cv}$).

It is a fundamental fact that this residual ``logarithmic'' coefficient of variation (\ref{CVlog}) is a decreasing function 
of the likelihood ratio between the truncated lognormal and the power law \cite{Castillo},
so, a ``large enough'' likelihood ratio corresponds
to a ``small enough'' $cv$
and this is what allows one to replace the likelihood ratio by $cv$ in the test
(which has the clear advantage that one avoids the maximum-likelihood estimation
of the parameters).
Note also that the distribution of $cv$ does not depend neither on the value of the exponent
nor on the value of $s_{cv}$;
it only depends on $n_{cv}$.

As the power law can be considered a particular instance of a truncated lognormal 
(one with $\mu -\ln s_{cv}\rightarrow -\infty$ and $\sigma^2\rightarrow \infty$,
with $\mu$ and $\sigma^2$ the mean and variance of the associated untruncated
normal distribution,
which leads to power-law exponent $\beta=1+|\mu-\ln s_{cv}|/\sigma^2$, see Refs. \cite{Castillo,Malevergne_Sornette_umpu}),
the likelihood ratio in this case will correspond to that of nested distributions, i.e., 
the power law is nested into the lognormal, which constitutes then a more general distribution than the former.
So, it should be clear that a truncated lognormal will fit a tail at least as well as a power law.
The point is if the improvement given by the lognormal is significant or not.
Note that this test constitutes the uniformly most powerful unbiased test for power law against truncated lognormality
\cite{Castillo,Malevergne_Sornette_umpu}.
Considering different values of $s_{cv}$ we will be able to determine if there is a transition between 
a power-law tail ($cv$ close to 1) and a lognormal ($cv$ far from 1) 
as the tail domain is increased; 
in other words, 
at which value of $s_{cv}$ a hypothetical power-law tail starts.
We refer to such a value, if it exists
(the value for which $s_{cv}$ crosses the critical line 
given by the percentile corresponding to the desired confidence level)
as $s_{pl}$.
More details are given in Ref. \cite{Corral_Gonzalez}.

\begin{figure}[ht]
\includegraphics[width=.80\columnwidth]{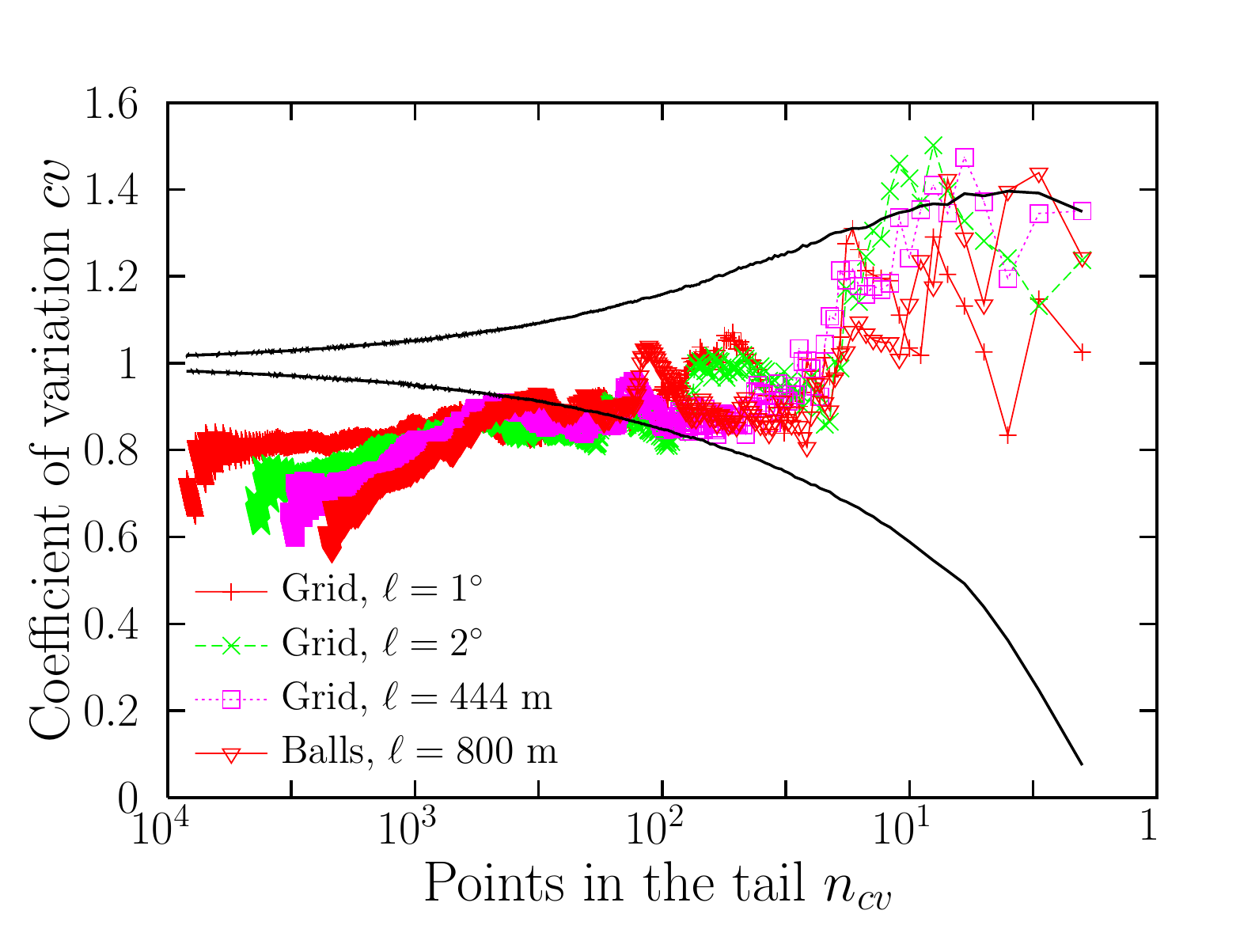}
\caption{
Empirical value of the ``residual logarithmic coefficient of variation'' $cv$ 
of the cluster population 
as a function of 
the number of residual points in the tail of the distribution, $n_{cv}$.
Several examples are shown, corresponding to our three approaches.
The solid lines correspond to
the simulated 0.05 and 0.95 percentiles of the distribution of $cv$
under the power-law null hypothesis.
In all cases the first crossing below the 0.05 percentile
takes place for $n_{cv}$ around 100,
which corresponds to $n_{pl}$ in Table \ref{table1new}.
Note that the horizontal axis is reversed, to display the tail on the right side. 
}
\label{fsonebyone}
\end{figure}

Table \ref{table1new} and Fig. \ref{fsonebyone} 
incorporate the results of this approach.
For the critical values of the test 
we take the percentiles 5 and 95 of the distribution of $cv$, 
which leads to a 90 \% confidence that the tail is power law,
and a 95 \% that the tail is power law in front of lognormal
($cv$ cannot be larger than one for the lognormal, so the test has to be one-sided).
We observe in the table that,
except for the largest considered cell width $\ell$, the cut-off value of the hypothetical
power-law tail $s_{pl}$ is in a range from 1500 to 10000 (inhabitants).
The number of population clusters (cities) covered by that range (number of points in the tail, $n_{pl}$)
turns out to be rather small, from 60 to 230, roughly.

We can use a variation of the logarithmic-coefficient-of-variation test
(in fact, its original linear form, essentially)
to rule out that the distribution of cluster size has an exponential tail,
as was claimed in other contexts \cite{Bernhardsson}
(and already criticized in Refs. \cite{Font-Clos2013,Corral_Font_Clos_PRE17}).
If we compute the usual residual coefficient of variation 
of the cluster size (just dividing the standard deviation and the mean of 
the difference between $s$ and a lower cut-off $s_{cv}'$, i.e.,
$s-s_{cv}'$)
and compare with the results expected for an exponential variable \cite{Castillo}, 
we get 
that $cv$ turns out to be above the 0.95 percentile,
which rules out the exponential tail
for any value of $s_{cv}'$.
Thus, the tail of the cluster-size distribution is not exponential.
Similar conclusions are reached if one uses as a test statistic the
mean of $s-s_{cv}'$ divided by its maximum \cite{Rochet_serra}.

\begin{widetext}
\begin{table}[ht]
\caption{
Population-cluster properties obtained from 
the grid-in-degrees, grid-in-km, and the ball approaches ($N$ and $s_{max}$);
results from the logarithmic-coefficient-of-variation test ($s_{pl}$ and $n_{pl}$),
together with a power-law fit ($\beta$);
and
results from the lognormal fit.
$N$ is total number of clusters,
$s_{max}$ is the size of the largest one 
(always associated to Barcelona),
$s_{pl}$ is the cut-off value for the power-law tail,
$n_{pl}$ is the number of clusters in that tail,
$s_{ln}$ and $n_{ln}$ are the equivalent for the lognormal fit,
$r$ is the number of orders of magnitude of that fit
($r=\log_{10}(s_{max}/s_{ln}$)),
$\mu$ and $\sigma$ are the selected lognormal parameters,
and $p$ is the $p-$value of the lognormal fit.
The last row for each approach corresponds to the aggregation for different $\ell$
of the rescaled variable $s/\langle s \rangle$.
The results for municipalities are also included, using data from IDESCAT at January 1, 2016.
The values of $s_{ln}$ swept are 50 per order of magnitude
and the $p-$value is computed from 1000 Monte-Carlo simulations.
}
\begin{tabular}{| r| rc | rr r| rr rrcc|}
\hline
$\ell$ & $N$ & $s_{max}$ & $s_{pl}$  & $n_{pl}$ & $\beta $ & $s_{ln}$ & $r$ & $n_{ln}$ & $\mu$ & $\sigma$ & $p$ \\
\hline 
$1^\circ$ &      8376 &   $1.98 \cdot 10^6$ &  6893  &    134  &  1.95   &  52.5 &      4.576 &      2450 &        2.120     &   3.102  &  0.21\\
$2^\circ$ &      4465 &   $2.48 \cdot 10^6$ &  10216  &    91   &   1.95  &  13.2 &      5.275 &      2571 &        3.138    &    2.975  &  0.23\\
$4^\circ$ &      2976 &   $3.05 \cdot 10^6$ &   13033 &      57   &  1.83   &  8.3 &      5.565 &      2082 &        3.643    &    2.869  &  0.40\\
$8^\circ$ &      1827 &   $4.06 \cdot 10^6$ &  1473  &    203     & 1.72  &  5.2 &      5.888 &      1470 &        3.933    &    2.814  &  0.25\\
$16^\circ$ &       694 &   $6.00 \cdot 10^6$ &   1462   &    101    &  1.69  & 2.1 &      6.458 &       635 &        4.665    &    2.653  &  0.33\\
$32^\circ$ &       102 &   $7.27 \cdot 10^6$ &    236   &     61   &  1.47  & 14.5 &      5.702 &        82 &        6.426     &   2.415  &  0.25\\
$1$ to $8^\circ$
&     17644 &   $2.18 \cdot 10^3$ & - & - & - & 0.02 &   5.059 &      7998 &       -4.735    &    3.064  &  0.44\\
\hline 
111 m&     10258 &   $1.94 \cdot 10^6$ &  7468     &  125   & 1.96 & 50.1 &      4.587 &      2810 &        1.542    &    3.176  &  0.23\\
222 m &      4966 &   $2.11 \cdot 10^6$ & 9763     &   92  &  1.90 & 10.5 &      5.304 &      2972 &        2.772     &   3.062  &  0.26\\
444 m &      3209 &   $3.05 \cdot 10^6$ &  6913    &    98  & 1.82  & 12.0 &      5.404 &      2047 &        3.604    &    2.866  &  0.21\\
889 m &      2095 &   $3.62 \cdot 10^6$ &  1462    &   228 & 1.72  &  6.3 &      5.758 &      1613 &        3.980    &    2.788  &  0.32\\
1778 m &       922 &   $6.00 \cdot 10^6$ &  1462    &   123  & 1.71 & 3.0 &      6.298 &       814 &        4.532    &    2.615  &  0.32\\
3556 m &       159 &   $7.18 \cdot 10^6$ & 635    &   70   & 1.59 & 22.9 &      5.496 &       126 &        6.598     &   2.131  &  0.27\\
$111$ to $889$ &     20528 &   $2.62 \cdot 10^3$ &  - & - & - & 0.03 &   4.998 &      8458 &       -4.843    &    3.076  &  0.27\\
\hline
100 m &     10263 &   $1.88 \cdot 10^6$ &  6731 & 135 & 1.95 & 26.3 &     4.853 &      3875 &         0.957   &     3.305  &  0.28\\
200 m &      5029 &   $2.41 \cdot 10^6$ & 9764  & 95 & 1.95 &14.5 &     5.222 &      2833 &        3.131   &     2.921  &  0.32\\
400 m &      3363 &   $2.66 \cdot 10^6$ & 11263 & 74 & 1.86 & 7.6 &     5.545 &      2358 &        3.466    &    2.937  &  0.23\\
800 m &      2262 &   $3.38 \cdot 10^6$ & 2736 & 163 & 1.74 & 6.3 &     5.729 &      1744 &        3.997    &    2.782  &  0.39\\
1600 m &      1059 &   $5.08 \cdot 10^6$ & 1492 & 148 & 1.71 & 3.0 &     6.226 &       941 &        4.578   &     2.640  &  0.30\\
3200 m &       220 &   $7.14 \cdot 10^6$ & 486 & 93 & 1.61 & 251.\phantom{0} &     4.454 &       129 &        3.959    &    2.824  &  0.27\\
100 to 800 &     20916 &   $2.54 \cdot 10^3$ & - & - & - & 0.02 &     5.085 &      9622 &       -4.749       & 3.051  &  0.21\\
\hline
 municip. &       947 &   $1.60 \cdot 10^6$ &  10870    &   110& 2.01  & 100 &    4.205 &       919 &        6.445    &    2.178  &  0.23\\
\hline
 \end{tabular}
 \label{table1new} 
\end{table} 
\end{widetext}

\subsection{Lognormal fits}

As the existence of a power-law tail does not rule out the existence of a lognormal tail,
and due to the low range of the power-law tail, 
and due also to the fact that we have ruled out the existence of a power law for small $s$,
as a next step we explore the performance of a lognormal fit.
Concretely, a lower-truncated lognormal (ln) distribution is given by a probability density
\begin{equation}
f_{ln}(s)=
{\sqrt{\frac 2\pi}}
\left[
 \mbox{erfc}\left(\frac{\ln s_{ln} -\mu}{\sqrt{2} \sigma}\right)
\right]^{-1}
\frac 1{ \sigma s}
\exp\left(-\frac{(\ln s-\mu)^2}{2\sigma^2}\right),
\label{Eqlognormal}
\end{equation}
defined for $s$ above the lower cut-off $s_{ln}$, 
with 
erfc the complementary error function,
and
$\mu$ and $\sigma$ the mean and standard deviation
of the associated untruncated normal distribution
($e^\mu$ turns out to be the scale parameter $\theta$ of $f_{ln}(s)$
and 
$\sigma$ its shape parameter).

We fit this truncated lognormal distribution to our population 
data extending to lognormals the method introduced 
in Refs. \cite{Peters_Deluca,Corral_Deluca} for (continuous) power laws, 
consisting in maximum-likelihood estimation 
plus Kolmogorov-Smirnov goodness-of-fit test \cite{Corral_Gonzalez}.
Although in recent years the recipe of Clauset et al. \cite{Clauset}
has become very popular for power-law tails, we prefer the more intuitive approach 
of Refs. \cite{Peters_Deluca,Corral_Deluca,Corral_Gonzalez} 
(the reasons for our choice derive in part from the results of 
Refs. \cite{Corral_nuclear,Voitalov_krioukov}
and also from the present research).
The extension of the method to lognormals has also been used in Ref. \cite{Corral_Gonzalez}.
At the end, we arrive to optimal values of the three parameters
$s_{ln}$, $\mu$, and $\sigma$, which are included in Table \ref{table1new}.

We see in the table how,
in contrast to the power-law tail, 
the lognormal fit covers a considerable range of data, 
with a rather small value of $s_{ln}$ (from 2 to 50 inhabitants, 
which leads to fits valid for more than 4 orders of magnitude in population)
and a relatively large $n_{ln}$ (either $n_{ln} \simeq N$ or $n_{ln}> 2000$).
Inadequacy of the lognormal to fit the smallest values of $s$
is expected due to the fact that the lognormal is a continuous distribution
and $s$ is a discrete variable.

\begin{figure}[ht]
\includegraphics[width=.90\columnwidth]{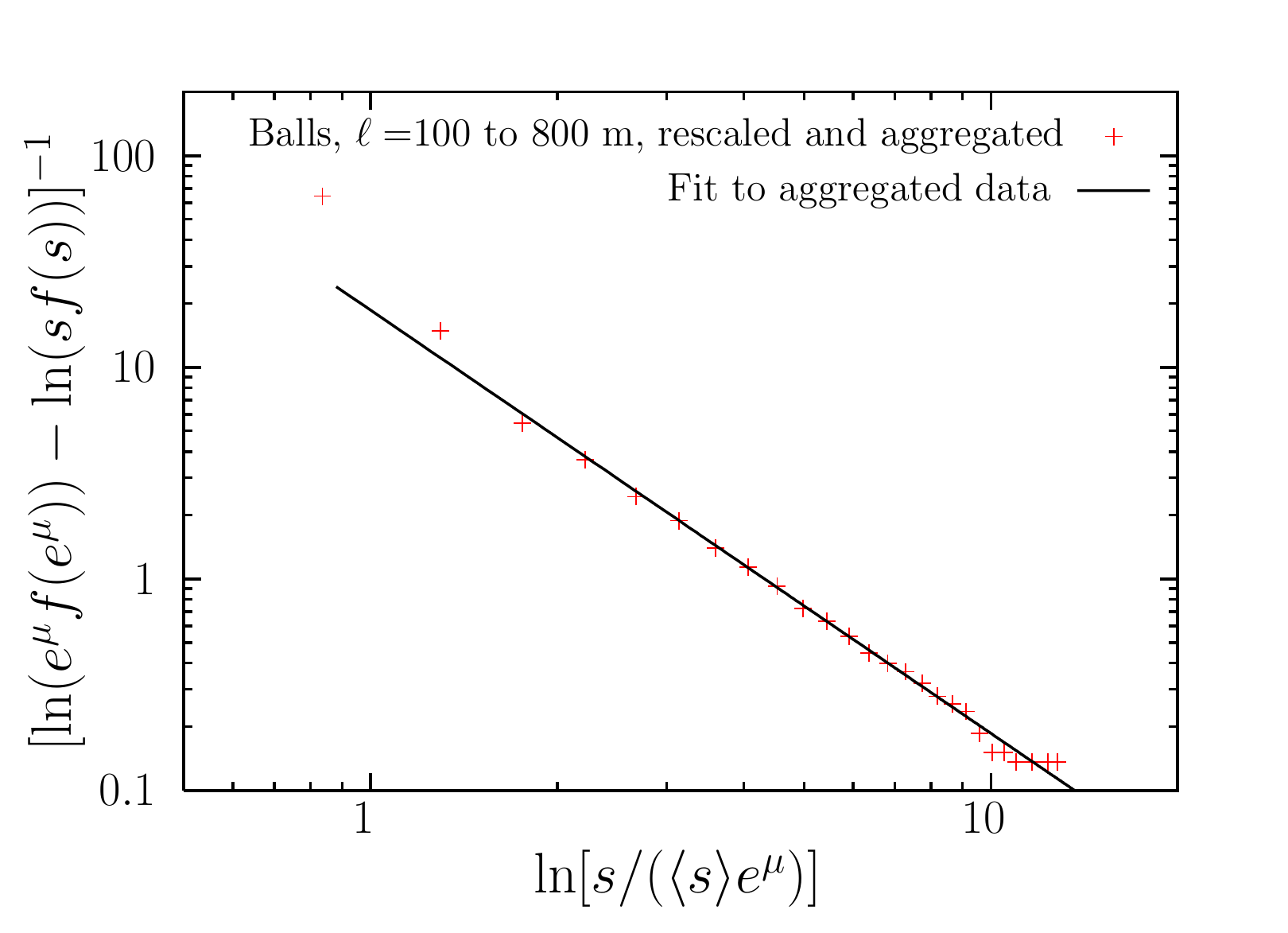}
\caption{
Empirical probability density of rescaled cluster sizes ($s/\langle s \rangle$)
aggregated for diverse values of $\ell$ (100, 200, 400, and 800 m),
in the ball approach.
The axes have been transformed to evaluate the goodness of the lognormal fit
(Table \ref{table1new}, continuous line).
Plot equivalent to Fig \ref{fsrescaled}(a), same data.
}
\label{lognormal_lineal}
\end{figure}

The performance of the fits can be visually appreciated from Fig. \ref{fs};
however, in order to stress the lognormal behavior, we apply a transformation
which should lead to a linear plot in the lognormal case, 
see Fig. \ref{lognormal_lineal}.
This consists in representing
$$
-\ln [\ln \left(e^\mu f_{ln}(e^\mu) \right) - \ln \left(s f(s) \right)]
\mbox{ versus }
\ln \ln \frac s {e^\mu}
$$
(restricted to $s > e^\mu$, 
to avoid the overlap with the branch $s < e^\mu$), 
or, equivalently, 
$[\ln \left(e^\mu f_{ln}(e^\mu) \right) - \ln \left(s f(s) \right)]^{-1}$
versus 
$\ln s - {\mu}$ in additional logaritmic scale on each axis.
Note that $f(s)$ refers to the empirical estimations of the density
whereas $f_{ln}(e^\mu)$ refers to the theoretical distribution evaluated at $s=e^\mu$
(so, both $\mu$ and $\sigma$ need to be estimated from data);
note also that 
both axes are doubly logarithmic.
Indeed, from Eq. (\ref{Eqlognormal}) we get
$$
-\ln \ln \left[ \frac{e^\mu f_{ln}(e^\mu)}{s f(s)}\right]
=  -2 \ln \ln \frac s {e^\mu} + \ln (2\sigma^2),
$$
which is a straight line with slope -2, 
in the variables defined above.
We see in Fig. \ref{lognormal_lineal} how
the straight behavior is more apparent than in the usual log-log plot of $f(s)$ versus $s$
(Fig. \ref{fs}), 
so we have an additional visual support for the lognormal fit
in front of the power law.
In fact, an additional shift by $2\ln\ln \langle s \rangle$
is applied in the figure, 
in order to collapse the different distributions, 
which are merged into a single one.

\subsection{Scaling to fix the lognormal parameters}

From Table \ref{table1new} one also realizes 
that the lognormal scale parameter $e^\mu$ increases with the grid or ball size $\ell$
but the shape parameter $\sigma$ keeps constant, roughly (from 2.5 to 3, 
except for the largest $\ell$).
We now take advantage of the fact that the cluster-size distributions
display scaling (at least approximately, see Fig. \ref{fsrescaled}a).
Therefore, for the different data sets (corresponding to different $\ell$) 
we can rescale $s$ as $s/\langle s \rangle$ (as in Fig. \ref{lognormal_lineal}), 
and then merge the different data sets into a single one, 
to which we can fit the lognormal distribution.
The results in Table \ref{table1new}
confirm that a single lognormal can fit the rescaled distributions
corresponding to different $\ell$.
Denoting the new value of the parameter $\mu$ as $\mu'$
(corresponding to the rescaled data), 
and taking into account the simple relation between
the original and the rescaled data,
we can express the value of $\mu$ of the original (not rescaled) distributions as
$\mu=\mu' + \ln \langle s \rangle$, 
which shows indeed that $\mu$ increases with $\ell$, 
as $\langle s \rangle$ increases with $\ell$.
Summarizing, and approximating the results from the table, we can write 
$e^\mu\simeq \langle s \rangle / e^{4.8}$ and $\sigma \simeq 3$.
For the dependence of $\langle s \rangle$ on $\ell$, 
using the data-collapse optimization method explained in Ref. \cite{Deluca_npg},
we find
$\langle s \rangle \propto \ell^{0.8}$;
nevertheless, the uncertainty of this exponent is rather large.

\subsection{Power-law fits and comparison with lognormal}

If one insists in fitting a power law to the tail of the distribution
(taking advantage of the fact 
that the tail of a lognormal becomes asymptotically a power law,
the tail being understood as the power-law tail defined by the 
logarithmic-coefficient-of-variation test as calculated above)
we find exponents roughly close to but below $\beta=2$ 
(the Zipf's value, see Table \ref{table1new});
nevertheless, for the reasons mentioned above, we consider the power-law fit
as anecdotic and prefer the lognormal fit, valid for a much larger range 
(Table \ref{table1new}).

In fact, we can make a quantitative comparison between power-law and lognormal fits.
For the power-law tail 
one already knows
that the lognormal fit has to yield a likelihood at least as large as the one given by the power law (when both are fitted over the same range), 
but this difference is not significant 
(this is what the $cv-$test is about, allowing one to define
the power-law tail in the range $s > s_{pl} $). 
In this sense, the power law always wins over the lognormal.
However, we are interested in fitting not just a short power-law tail, 
but as much as possible of the bulk of the distribution (including the tail).
So, on the one hand we have (model 1)
the lognormal fit, in the range $s \ge s_{ln}$, 
and on the other hand (model 2) we have to consider the power-law tail 
complemented by the lognormal fit from $s_{ln}$ to $s_{pl}$.
Note that model 2 yields 4 parameters 
($\mu$ and $\sigma$ plus $s_{pl}$ and $\beta$, considering $s_{ln}$ as fixed); 
whereas model 1, the single lognormal, only has 2 parameters
($\mu$ and $\sigma$).
In mathematical terms, the two models can be expressed as
$f_1(s)=f_{ln}(s)$ for $s \ge s_{ln}$ and
$$
f_2(s)=\left\{
\begin{array}{ll} 
f_{ln}(s)                          & \mbox{ for } s_{ln} \le s \le s_{pl}\\
f_{pl}(s) n_{pl}/ n_{ln}    & \mbox{ for } s > s_{pl},
\end{array} 
\right.
$$
where the factor $n_{pl}/n_{ln}$ ensures normalization
(note that we do not impose continuity of $f_2(s)$ at $s=s_{pl}$,
that would reduce the likelihood of the resulting fit).

We can compute the difference in log-likelihoods of both models as
$$\Delta \ln \hat L = \ln \hat L_2 - \ln \hat L_1 = 
\sum_{i=1}^{n_{ln}} \left(\ln f_2(s_i) -\ln f_1(s_i)\right),
$$
(we only number the clusters with $s\ge s_{ln}$, and in increasing order,
$s_1\le s_2 \le  \dots \le s_{n_{ln}}$);
however,
as the two models coincide in the range $s_{ln} \le s < s_{pl}$,
the comparison of likelihoods only needs to be done at the tail, 
$s\ge s_{pl}$, and thus,
$$\Delta \ln \hat L = 
\sum_{i={n_{ln}-n_{pl}+1}}^{n_{ln}} 
\left(\ln f_{pl}(s_i;\beta,s_{pl}) -\ln f_{ln}(s_i;\mu,\sigma,s_{pl})\right),
$$
where we have made explicit the dependence on the parameters.
Note that for the lognormal we have replaced its lower cut-off $s_{ln}$
by $s_{pl}$, this is what allows to eliminate the factor $n_{pl}/n_{ln}$
that multiplied $f_{pl}(s)$ as in this way both distributions are normalized 
in the range $s\ge s_{pl}$.

As expected, this difference of log-likelihoods turns out to be negative 
(see Table \ref{table2}),
which means that model 1 (single lognormal) would be favored;
however, this comparison does not take into account the different number of parameters.
If we introduce the Akaike information criterion, 
$AIC=2 k - 2 \ln \hat L$, where $k$ is the number of parameters, 
we get $\Delta AIC =4 - 2 \Delta \ln\hat L$, which is positive, favoring more clearly
the lognormal model (Table \ref{table2} again).
The Bayesian information criterion, 
$BIC = k \ln n_{ln} -2 \ln \hat L$ leads to 
$\Delta BIC =2 \ln n_{ln}  - 2 \Delta \ln\hat L$,
with the same conclusion.
In any case, the lognormal fit is preferred.

Note that when only the tail is compared, the power law has one parameter
whereas the lognormal has two, which favors the former;
however, considering the whole range $s \ge s_{ln}$,
the situation is reversed, as
the power-law tail combined with the lognormal bulk has four parameters, 
which favors the single lognormal, despite the fact that the difference in log-likelihood
does not change.  
Note also that the difference in $AIC$ does not depend on the number of data in the fit,
whereas $\Delta BIC$ does; 
in fact, the more data, the better the simple (lognormal) fit.

Finally, in order to allow some comparison, 
we apply the same test as in Ref. \cite{Levy_comment}, 
which is Pearson's chi-squared test.
We consider just two classes, $s_{ln} \le s < s_{pl}$ and $s \ge s_{pl} $,
and compute $x^2=\sum_{k=1}^2 (E_k-O_k)^2/E_k$, 
where $O_k$ is the observed number of clusters in the $k-$th class
(either $n_{ln}-n_{pl}$ or $n_{pl}$) and $E_k$ is the expected number of clusters
from the lognormal fit, which is either $(1-q_{tail})n_{ln}$ or $q_{tail} n_{ln}$,
with $q_{tail}$ the probability that the lognormal fit assigns to the tail.
This is calculated as 
$$
q_{tail}=
 \mbox{erfc}\left(\frac{\ln s_{pl} -\mu}{\sqrt{2} \sigma}\right)
\left/
 \mbox{erfc}\left(\frac{\ln s_{ln} -\mu}{\sqrt{2} \sigma}\right).
\right.
$$
The results, included in Table \ref{table2} show that the values of $x^2$ are in all cases too small 
to reject the lognormal fit.

\begin{widetext}
\begin{table}[ht]
\caption{
Model comparison between single lognormal fit (model 1, with 2 parameters)
and lognormal bulk plus power-law tail (model 2, with 4 parameters).
Differences are computed as model 2 minus model 1.
The outcome always favors the simpler model 1, lognormal. 
Results for the chi-squared test explained in the text 
(see also \cite{Levy_comment}) applied to model 1 
are also included
and confirm that the lognormal fit cannot be rejected.
}
\begin{tabular}{| r| ccr| c|}
\hline
$\ell$ & $\Delta \hat L$ & $\Delta AIC$ & $\Delta BIC $ & $\chi^2$ model 1 \\
\hline 
$1^\circ$ &       -1.269    &     6.538     &   12.334 & 0.0012\\
$2^\circ$ &     -1.105     &    6.209     &   11.231  & 0.0009\\
$4^\circ$  &     -0.632      &   5.264     &    9.350  & 0.4684\\
$8^\circ$  &     -1.381      &   6.761     &   13.388 & 0.8624\\ 
$16^\circ$  &     -1.603      &   7.207     &   12.437 & 0.9272\\ 
$32^\circ$  &     -0.610      &   5.220     &    9.441  & 0.8631\\
\hline
111 m  &     -1.152      &   6.305     &   11.961 &  0.0006\\ 
222 m  &     -1.380      &   6.760     &   11.804 & 0.2616\\ 
444 m  &     -1.158      &   6.316     &   11.486 & 0.6692\\ 
889 m  &     -1.446      &   6.893     &   13.752 & 1.2255\\ 
1778 m  &     -2.179      &   8.358     &   13.982 & 0.6197\\ 
3556 m  &     -1.221      &   6.441     &   10.938 & 0.0000\\ 
\hline
100 m  &     -1.030      &   6.060     &   11.870 & 0.1532\\ 
200 m  &     -1.413      &   6.826     &   11.934 & 0.0147 \\ 
400 m  &     -1.346      &   6.692     &   11.300 & 0.2874\\ 
800 m  &     -1.364      &   6.728     &   12.916 & 1.3501\\ 
1600 m  &     -1.492      &   6.983     &   12.978 & 0.5327\\ 
3200 m  &     -1.658      &   7.317     &   12.382 & 0.3306\\ 
\hline
 municip. 
  &     -1.559      &   7.118     &   12.519 & 0.0054\\ 
\hline  
 \end{tabular}
 \label{table2} 
\end{table} 
\end{widetext}


\subsection{Dragon-king effect}

It is remarkable that, 
although the cluster associated to the city of Barcelona
could be considered to have the status of dragon king 
(in the Sornette's sense of a very large outlier in $s$ \cite{Sornette_dragon_king}), 
with a population much larger than that
of the second largest cluster (at least a 5-fold larger, depending on $\ell$),
with a clear deviation therefore with respect Zipf's law in terms of the rank-size representation,
this does not cause the rejection
neither of the lognormal fit nor of the power-law tail
for the cluster-size distribution.
In other words, in terms of the distribution of sizes,
the fits are not affected by the dragon-king effect.

We can go one step further and study the influence of the largest cluster on
the distribution, just removing it. 
The results,
show a very similar behavior, except that the value of the exponent $\beta$ gets somewhat larger.
We conclude that, 
from the point of view of the distribution of cluster sizes,
and for the present study, 
the largest cluster 
(which could be considered a dragon king) 
does not change the character of the tail, 
and only modifies a little the value of the parameters.

\subsection{Origin of the large variability in city sizes}

Our approach allows us to investigate the origin of the broadness of the 
distribution of the size of population clusters
(i.e., the size of cities, in our definition).
In the grid approach,
the distribution $f(s)$ arises from the sum of the number of inhabitants 
$h$ of each cell
(with a distribution given by $f(h)$) along each cluster, 
so, $f(s)$ depends both on $f(h)$ and on the distribution of the number of cells per cluster;
nevertheless, this is not enough, as one needs to take into account that 
the values of $h$ are not independent from cell to cell,
i.e., there are spatial correlations in the values of $h$
(highly populated cells tend to be surrounded by highly populated cells, 
and reciprocally).
This is in fact an obvious fact, but we can demonstrate the relevance of correlations in the values of $h$ eliminating these correlations 
and looking for the resulting $f(s)$.

\begin{figure}[ht]
\includegraphics[width=.80\columnwidth]{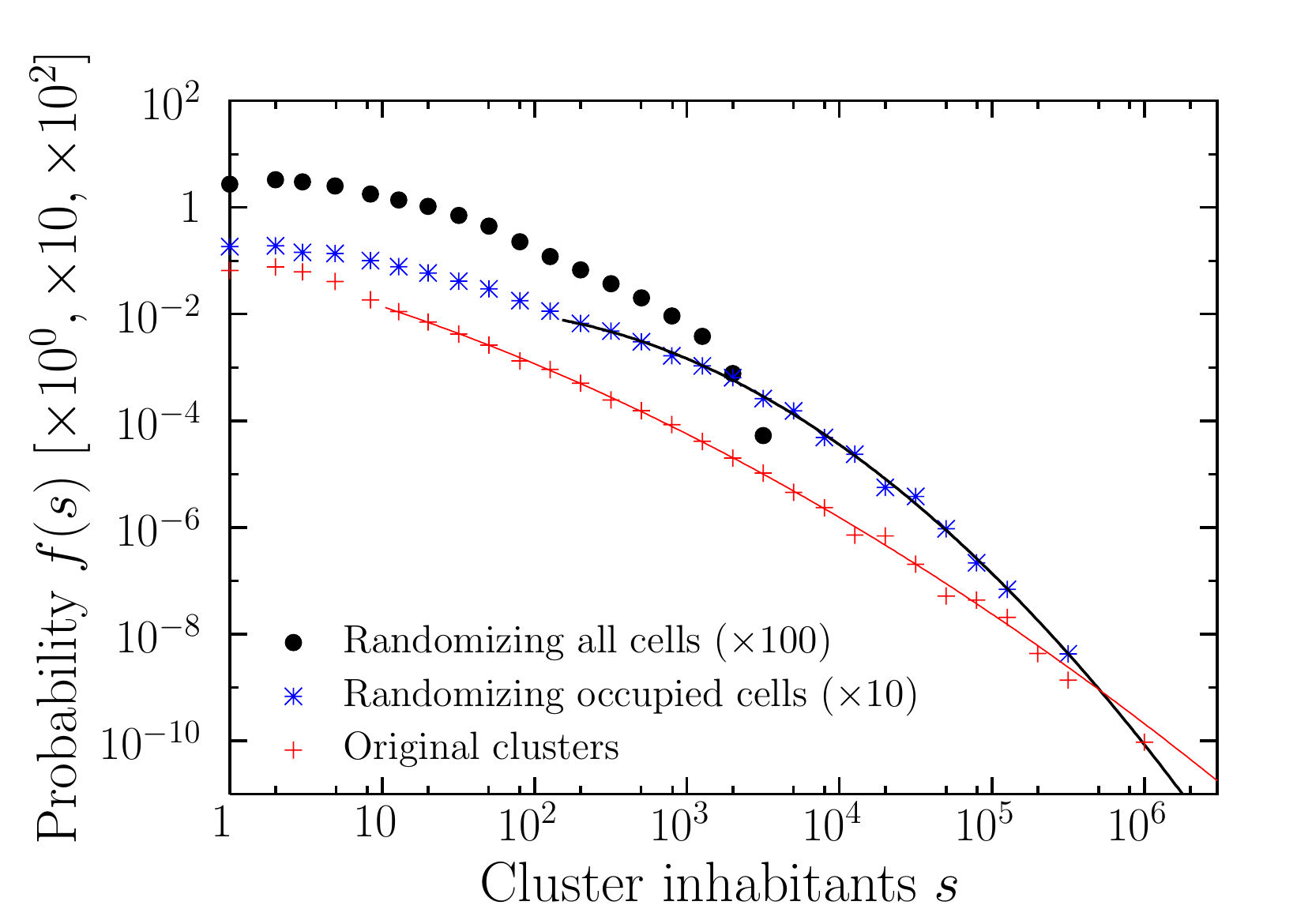}
\caption{
Comparison of the original distribution of cluster sizes $f(s)$
with the distribution of cluster sizes obtained when the values $h$ of
the population of each cell are randomized between occupied cells.
The less interesting case 
in which the values of $h$ are randomized between all cells
(occupied and unoccupied) is also shown. The example shown
corresponds to the grid-in-km approach, with $\ell=222$ m.
}
\label{Frandomize}
\end{figure}

We eliminate the correlations just reshuffling the values of $h$ among occupied clusters; 
this keeps the distribution $f(h)$ and the spatial extend of clusters unchanged.
The results are displayed in Fig. \ref{Frandomize}, 
showing that the resulting randomized $f(s)$ is less broad than
the original $f(s)$; in particular, 
no sign of an approximate power-law tail with exponent close to 2 is found,
leading to the conclusion that the main cause of the large variability in city size
(and the cause of the rough Zipf-like behavior) are spatial correlations.
This means that the (fractal) shape of cities is not enough to explain their population
distribution.
Above we mentioned the advantages of the ball approach over the grid approach;
however, notice that 
this reshuffling procedure can only be performed under the grid approach,
so, both approaches can be considered as complementary.

\section{Conclusions}

Population data of high spatial resolution allows one to locate individuals
and to build clusters of them in space.
If the locations of individuals correspond to their residence place
(as it is in our study),
these clusters constitute a natural definition of human settlements
or, broadly speaking, cities.
We have scrutinized the distribution of the number of individuals
in the clusters obtained in this way (natural-city size)
with up-to-date data-analysis tools.
On the one hand, scaling analysis allows us to rule out the existence
of a power-law size distribution with exponent $\alpha > 1$
for the smallest cities;
on the other hand, the logarithmic-coefficient-of-variation test
shows that a power-law tail has a very limited range of applicability.
Instead, a lognormal fit holds for a considerable part of the data
and does not only yield a higher likelihood
than a model joining a lognormal part in the bulk plus a power-law tail,
but also has less parameters, 
which makes the lognormal to be clearly supported by model comparison 
using AIC or BIC.

Obviously, we do not dispute that
the US population distribution, as measured from its census, 
is better described by a power law than by a lognormal at its tail,
as claimed in other studies
\cite{Levy_comment,Malevergne_Sornette_umpu}.
But it could be that the bad performance of the lognormal at the tail
of the US city distributions is due to the fact that
it is a pure (untruncated) lognormal which is fitted, 
and not a truncated one, as used here.
In any case, our example sets clearly that other datasets
and/or other definition of cities
can lead to a better characterization by (truncated) lognormal distributions.
This implies that universality \cite{Stanley_rmp}
does not seem to be a characteristic of city-size distributions,
as also found for other complex systems, 
for instance, wildfires \cite{Hantson_Pueyo16,Corral_Gonzalez}.

Our spatial-grid based approach also allows us to stress 
the importance of spatial correlations in the broadness of the 
city-size distribution;
in other words, knowledge of the area occupied by cities
together with the distribution of individuals in small grids 
does not allow to explain the number of inhabitants in cities.
Finally, we have seen how, although our largest cluster
(associated to the city of Barcelona) has the status of a dragon-king,
it does not have an important influence on the city-size distribution.
The reason is that, under the distribution-of-sizes representation
(in contrast to the rank-size approach),
the largest cluster counts just as one single realization of the random variable
(in contrast to more than one million counts of individuals from that cluster in the other approach),
and thus, it has a very small statistical weight.

On a further step, one could 
fit to the data a power law truncated also from above
(a power law defined in a range $s_{pl1} \le s \le s_{pl2}$,
where these parameters are optimized by the fitting \cite{Corral_Deluca,Corral_Gonzalez}). 
This leads to exponents $\alpha$ in the range 1.6 to 1.8, 
valid for several orders of magnitude, between 3 and 4,
starting at values of $s$ around 500 or higher. 
In some cases, the upper truncation point turns out to be above $s_{max}$
(the largest cluster), which means that the fitting method
is not able to distinguish the power-law tail
(which has a somewhat different power-law exponent $\beta$). 
In any case, the lognormal fit 
turns out to be valid in a wider range, and it is thus preferred. 
In a future work one could compare the performace of the lognormal
with that of a double power law \cite{Corral_Gonzalez},
which has been proposed for other Zipf-like systems 
\cite{Ferrer2000a,Gerlach_Altmann}.

\section{Acknowledgements}

We are indebted to 
E. Su\~n\'e from IDESCAT,
S. Manrubia, J. del Castillo, and I. Serra.
A. C. has participated in the grant from the Spanish
MINECO awarded to the Barcelona Graduate School of Mathematics (BGSMath) under the Mar\'{\i}a de Maeztu Program (grant MDM-2014-0445),
as well as in the projects FIS2012-31324, FIS2015-71851-P (MINECO), and FIS-PGC2018-099629-B-I00 (MICIU)
and in 2014SGR-1307 (AGAUR). 
F.U. and E.A. 
acknowledge the funding of
PGC2018-101643-B-I00 (MICIU)
and
the EPSRC grant EP/M023583/1, 
respectively.


\end{document}